\documentclass{aa}
\usepackage{epsfig}

\catcode`\@=11
\def\gsim{\ifmmode{\mathrel{\mathpalette\@versim>}}
    \else{$\mathrel{\mathpalette\@versim>}$}\fi}
\def\lsim{\ifmmode{\mathrel{\mathpalette\@versim<}}
    \else{$\mathrel{\mathpalette\@versim<}$}\fi}
\def\@versim#1#2{\lower 2.9truept \vbox{\baselineskip 0pt \lineskip 
    0.5truept \ialign{$\m@th#1\hfil##\hfil$\crcr#2\crcr\sim\crcr}}}
\catcode`\@=12

\def\msun{M_{\odot}}
\def\Mg{M_{\rm g}}
\def\Mgn{M_{\rm g,11}}
\def\Mc{M_{\rm c}}
\def \M {{\cal M}}

\def\rhog{\rho_{\rm g}}
\def\rhoc{\rho_{\rm c}}
\def\rhocm{\bar{\rhoc}}
\def\rhocz{\rho_{\rm c0}}
\def\rhon{\rho_{\rm n}}
\def\rhot{\tilde\rho}

\def\Reff{R_{\rm e}}

\def\alphau{\alpha_1}
\def\alphad{\alpha_2}
\def\alphat{\alpha_3}
\def\alphai{\alpha_i}
\def\alphaj{\alpha_j}
\def\alphak{\alpha_k}
\def\alphaun{\alpha_{1,1}}

\def\acu{a_1}
\def\acd{a_2}
\def\act{a_3}
\def\aci{a_i}

\def\acun{a_{1,250}}

\def\tacd{\tilde{a}_2}
\def\tact{\tilde{a}_3}

\def\xv{{\bf x}}
\def\csiv{\mbox{\boldmath$\xi$}}
\def\tcsiv{\tilde{\csiv}}

\def\xci{x_i}
\def\xcj{x_j}
\def\csici{\xi_i}

\def\rt{\tilde r}
\def\yt{\tilde y}
\def\zt{\tilde z}
\def\rv{{\bf r}}

\def\nuv{\mbox{\boldmath$\nu$}}
\def\tnuv{\tilde{\nuv}}
\def\dcsiv{\dot{\csiv}}

\def\ddcsiv{\ddot{\csiv}}

\def\ddxv{\ddot{\xv}}
\def\ddrv{\ddot\rv}

\def\DePsi{\Delta\Psi}
\def\DePsit{\widetilde{\DePsi}}
\def\Dtau{\Delta (\tau)}
\def\phig{\phi_{\rm g}}

\def\tphig{\tilde{\phig}}

\def\phiM{\varphi_{\rm M}}
\def\theM{\vartheta_{\rm M}}
\def\psiM{\psi_{\rm M}}

\def\Omegv{{\bf \Omega}}
\def\dOmegv{\dot{\Omegv}}
\def\tOmegv{\tilde{\Omegv}}
\def\dtOmegv{\dot{\tOmegv}}

\def\Omegvc{\Omegv_{\rm c}}
\def\Omegvp{\Omegv'}

\def\tOmegvp{\tilde{\Omegvp}}

\def\omegphi{\omega_{\varphi}}
\def\omegthe{\omega_{\vartheta}}
\def\omegpsi{\omega_{\psi}}

\def\omegi{\omega_i}
\def\tomegi{\tilde{\omega}_i}
\def\omegapm{\omega_{\pm}}
\def\omegap{\omega_+}
\def\omegam{\omega_-}
\def\lambdapm{\lambda_{\pm}}
\def\rop{r_{+}}
\def\rom{r_{-}}

\def\dphi{\dot\varphi}
\def\dtheta{\dot\vartheta}
\def\dpsi{\dot\psi}

\def\ddphi{\ddot\varphi}
\def\ddtheta{\ddot\vartheta}
\def\ddpsi{\ddot\psi}

\def\omegac{\omega_{\rm c}}
\def\Omegac{\Omega_{\rm c}}

\def\Rot{{\cal R}}
\def\Rotp{\Rot'}
\def\Rotc{\Rot_{\rm c}}
\def\RotT{\Rot^{\rm T}}
\def\RotpT{\Rotp^{\rm T}}

\def\sphi{\sin\varphi}
\def\sthe{\sin\vartheta}
\def\spsi{\sin\psi}
\def\cphi{\cos\varphi}
\def\cthe{\cos\vartheta}
\def\cpsi{\cos\psi}

\def\Cc{C_{\rm c}}

\def\Tc{{\bf T}}
\def\Ti{T_i}
\def\Tj{T_j}
\def\Tu{T_1}
\def\Td{T_2}
\def\Tt{T_3}
\def\Tij{T_{ij}}
\def\DT{\Delta T}
\def\DTij{\DT_{ij}}
\def\DTdu{\DT_{21}}
\def\DTtu{\DT_{31}}
\def\DTtd{\DT_{32}}
\def\DTud{\DT_{12}}
\def\DTut{\DT_{13}}
\def\DTdt{\DT_{23}}

\def\wci{w_i}
\def\wcu{w_1}
\def\wcd{w_2}
\def\wct{w_3}

\def\Ii{I_i}
\def\Iu{I_1}
\def\Id{I_2}
\def\It{I_3}
\def\DI{\Delta I}
\def\DIdu{\DI_{21}}
\def\DItu{\DI_{31}}
\def\DItd{\DI_{32}}

\def\hg{h_{\rm g}}

\def\Porb{P_{\rm orb}}
\def\Pdyn{P_{\rm dyn}}

\def\Pphi{P_{\varphi}}
\def\Pthe{P_{\vartheta}}
\def\Ppsi{P_{\psi}}

\def\tH{t_{\rm H}}

\def\sigv{\sigma_{\rm V}}
\def\sigvn{\sigma_{\rm V,1000}}
\def\sigvd{\sigv^2}

\def\mmax{m_{\rm max}}
\def\mi{m_{\rm i}}

\def\Ntot{N_{\rm tot}}
 
\begin{document}

   \title{Collisionless evaporation from cluster elliptical galaxies:}

    \subtitle{a contributor to the intracluster stellar population}

    \author{V. Muccione\inst{1}
	    \and
            L. Ciotti\inst{2}}

    \offprints{V. Muccione}

    \institute{Observatoire de Geneve, ch. des Maillettes 51, 
	       1290 Sauverny, Suisse\\
	       email: veruska.muccione@obs.unige.ch
               \and
               Dipartimento di Astronomia, Universit\`a di Bologna, 
               via Ranzani 1, I-40127 Bologna, Italy\\
       	       email: luca.ciotti@unibo.it}
               
    \date{Accepted, April 3, 2004}

\abstract{By means of simple numerical models we discuss whether {\it
collisionless stellar evaporation} from cluster elliptical galaxies
could be an effective mechanism for the production of intracluster
stellar populations.  The effectiveness of this mechanism is due to
the fact that, for realistic galaxy and cluster models, the galaxy
oscillation periods near equilibrium configurations in the cluster
tidal field are of the same order of stellar orbital times in the
external parts of the galaxies themselves.  With the aid of
Monte-Carlo simulations we explore the evolution of stellar orbits in
oscillating galaxies placed near different equilibrium positions.  We
found that, over an Hubble time, the main effect is a substantial
expansion of the galactic outskirts, particularly affecting the galaxy
at the cluster center and those orbiting near the cluster core radius:
overall, approximately the 10\% of the galaxy mass is affected.  Thus,
the proposed mechanism could be of some importance in the shaping of
the halo of cD galaxies and in making easier ``galaxy harassment'' in
the formation of the intracluster stellar population.
\keywords{galaxies: elliptical and lenticular, cD -- galaxies: evolution --
galaxies: interactions -- galaxies: kinematics and dynamics}}
\authorrunning{Muccione and Ciotti}
\titlerunning{Origin of the intracluster stellar population}
\maketitle 

\section{Introduction}

Observational evidences of an intracluster stellar population
(hereafter ISP) are mainly based on the detection of planetary nebulae
and red giant branch stars freely floating in the intergalactic space.
For example, Theuns \& Warren (1996) identified 10 intergalactic
planetary nebulae in the Fornax cluster, while M\'endez et al. (1997)
detected 11 intergalactic objects in the Virgo cluster whose
cumulative luminosity functions is in good agreement with planetary
nebula luminosity function. In addition, Ferguson et al. (1998)
identified an intergalactic red giant branch stars population in the
Virgo cluster, while Feldmeier et al. (1998) (with observations of
three blank fields in the Virgo cluster), confirmed that a significant
fraction of Virgo's starlight is due to the ISP.  More recently
Okamura et al. (2002) have identified 38 candidates of intracluster
planetary nebulae in the core of the Virgo cluster.  Overall, the data
suggest that approximately 10\% (or even more) of the stellar mass of
the cluster is in intergalactic stars (e.g., see Ferguson et al. 1998,
Arnaboldi et al. 2002, Durrel et al. 2002, Arnaboldi et al. 2003,
Totani 2003, see also Arnaboldi, Gerhard \& Freeman 2003, and
references therein).

The usual scenario assumed to explain the finding above is that the
tidal interactions between galaxies (for example during a fast
encounter, e.g., see Merritt 1983, 1984, 1985), and of galaxies with
the cluster gravitational field, lead to a substantial stripping of
stars from galaxies to the parent cluster (the so-called ``galaxy
harassment'' scenario, see, e.g., Moore et al. 1996; Napolitano et
al. 2003).  In this paper we explore an additional ``stripping''
mechanism, namely we quantitatively discuss the effect of resonant
interaction between stellar orbits inside the galaxies and the cluster
tidal field (hereafter, CTF). The present study is supported by the
fact that the characteristic oscillation times of a galaxy near its
equilibrium position in the CTF and the mean stellar orbital times in
the galaxy external regions are of the same order of magnitude, as
already recognized by Ciotti \& Giampieri (1998, hereafter CG98).

In fact, Hawley \& Peebles (1975) reported a possible indication
(confirmed by Thompson 1976), that the galaxies are preferentially
aligned along the radius vector to the center of the cluster for the
Coma cluster; they also suggested the CTF as the possible cause of the
alignment.  The best evidence for alignment is for brightest cluster
galaxies: Adams, Strom \& Strom (1980) found, in 7 very elongated
clusters, a general trend for ellipticals to be aligned with the
cluster major axis, and also Trevese, Cirmele \& Flin (1992) found a
strong alignment of the brightest galaxy major with the long axis of
the parent cluster.  Numerical N-body simulations (Ciotti \& Dutta
1994, hereafter CD94) indeed confirmed the hypothesis that the CTF
could be at the origin of the observed alignment and, in particular,
revealed that their model elliptical galaxies (Es) behaved with good
approximation as rigid bodies. In a complementary analytical
exploration of this problem CG98 determined the equilibrium positions
of triaxial ellipsoids in the CTF for various cases, and showed that
the oscillations period of the galaxies, when slightly displaced from
their equilibrium configurations, are of the same order of magnitude
of the stellar orbital periods in the galaxy outskirts. This curious
finding naturally leads to ask what could be the effect of possible
``resonances'' between stellar orbital periods in cluster Es and their
oscillation periods. Unfortunately, the $N$-body simulations of CD94
were characterized by a limited number of particles (i.e., $N\simeq
3\times 10^4$) and by the usual softening, so that evaporations rates
could not be properly investigated.

In order to investigate the scenario described above, Muccione \&
Ciotti (2003a,b, hereafter MC03a,b) performed a few preliminary
Monte-Carlo simulations of orbital exploration of oscillating
galaxies, and the results were encouraging, with estimated escape
rates summing up to 10 per cent of the initial galaxy mass over an
Hubble time. In the present paper we considerably extend these two
previous investigations, by assuming that the galaxies are triaxial
ellipsoids and two cases are considered: when the center of mass of
the ellipsoid is at rest at the equilibrium point of the field
generated by the cluster, and when it is placed on a circular orbit
around the center of a spherically symmetric cluster.  After deriving
the equations of the motion of a star inside an oscillating galaxy, we
integrate numerically these equations under different initial
conditions realized by using a Monte-Carlo extraction.

The paper is organized as follows.  In Sect. 2 we briefly review the
proper physical setting of the problem, and in Sect. 3 we describe the
adopted galaxy and cluster models.  In Sect. 4 we describe in the
numerical integration scheme, while in Sect. 5 we present the main
results, that are finally summarized and discussed in Sect.  6. In the
Appendix, the specific technique adopted to obtain the gravitational
field inside the galaxy models and other useful dynamical quantities
are briefly described.

\section{The physical setting}

\subsection{Triaxial galaxy at the center of a triaxial cluster}

Following the analytical treatment of CG98, we start presenting the
simple case of a spinless galaxy with its center of mass at rest at
the center of a triaxial galaxy cluster. In CG98 it was shown that the
galactic equilibrium configurations correspond to the galaxy inertia
ellipsoid oriented along the CTF principal directions: without loss of
generality we assume that in the (inertial) Cartesian coordinate
system $C$, the CTF tensor $\Tc$ is in diagonal form, with components
$\Ti$ $(i=1,2,3)$.

By using three successive counterclockwise rotations ($\varphi$ around
$x$ axis, $\vartheta$ around $y'$ axis and $\psi$ around $z''$ axis),
CG98 showed that the linearized equations of motion for the galaxy
near the equilibrium configurations can be written as
\begin{equation}
\cases{ 
\ddphi   = \displaystyle{\DTtd\DItd\over\Iu}\varphi,\cr 
\ddtheta = \displaystyle{\DTtu\DItu\over\Id}\vartheta,\cr 
\ddpsi   = \displaystyle{\DTdu\DIdu\over\It}\psi,\cr}
\end{equation}
where $\DTij \equiv \Ti-\Tj$, and $\Ii$ are the principal components
of the galaxy inertia tensor.  If we also assume that
$\Tu\geq\Td\geq\Tt$ and $\Iu\leq\Id\leq\It$, then the equilibrium
position of eq. (1) is {\it stable} (for the geometrical
interpretation of this condition see Sect. 3), and the solution of
eq. (1) is
\begin{equation}
\varphi   =\phiM \cos (\omegphi t),\;
\vartheta =\theM \cos (\omegthe t),\; 
\psi      =\psiM \cos (\omegpsi t),
\end{equation}
where
\begin{equation}
\cases{
\omegphi = \displaystyle\sqrt{\DTdt\DItd\over\Iu},\cr
\omegthe = \displaystyle\sqrt{\DTut\DItu\over\Id},\cr 
\omegpsi = \displaystyle\sqrt{\DTud\DIdu\over\It},\cr}
\end{equation} 
are the three independent oscillations frequencies. For simplicity we
assume that at $t = 0$ the galaxy is at the maximum deviation from the
equilibrium configuration (with null angular velocity).

For computational reasons the best reference system in which calculate
and discuss the stellar orbits is the oscillating reference system
$C'$ in which the galaxy is at rest, and its inertia tensor is in
diagonal form. As a consequence of this choice, the equation of the
motion for a galactic star in $C'$ is
\begin{equation}
\ddcsiv = \RotT\ddxv 
        -2\Omegv\wedge\nuv 
        -\dOmegv\wedge\csiv 
        -\Omegv\wedge (\Omegv\wedge\csiv ),
\end{equation}
where $\csiv$ and $\nuv=\dcsiv$ are the stellar position and velocity
in $C'$, the suffix ``T'' means transpose, and $\Rot$ is the
orthogonal transformation matrix between $C$ and $C'$ so that
$\xv=\Rot\csiv$.  The explicit form of $\Rot$ is given by eq. (9) in
CG98: as well known, this matrix define the angular velocity of $C'$
that, expressed in $C'$, reads
\begin{eqnarray} 
\Omegv &=& (\dphi\cthe\cpsi +\dtheta\spsi ,\nonumber\\ 
       && -\dphi\cthe\spsi +\dtheta\cpsi ,
          \dphi\sthe      +\dpsi).
\end{eqnarray}
If we indicate with $\phig = \phig (\csiv)$ the galactic gravitational
potential in $C'$, then in eq. (4)
\begin{equation}
\RotT\ddxv\equiv -\nabla_{\csiv}\phig +\RotT\Tc\Rot\csiv,
\end{equation}
where $\nabla_{\csiv}$ is the gradient operator in $C'$, and the
direct effect of the cluster gravitational field is obtained in the
tidal approximation. In this scheme, each star experience two
different influences from the ambient cluster: an {\it indirect} one,
due to the induced oscillatory motion of the galaxy, and the {\it
direct} acceleration due to the galaxy and the cluster mass
distributions. Of these two last terms, the by far more important
within the galaxy is the galactic gravitational field. In Sects. 3 and
4, and in Appendices B and C, we give the explicit expression for the
quantities appearing in eqs. (4) and (6) for the adopted galaxy and
cluster models.

\subsection{Triaxial galaxy on circular orbit in a spherical cluster}

The second, more complicate (but still highly idealized, although more
astrophysically relevant) case of interest is represented by a
triaxial galaxy with its center of mass in circular orbit of radius
$r$ in a plane (say, $z=0$) of a spherical cluster, with angular
velocity $\Omegvc =(0,0,\Omegac)$, where
\begin{equation}
\Omegac^2(r)={G\Mc (r)\over r^3},
\end{equation} 
and where $\Mc(r)$ is the cluster mass within $r$. Following the
treatment of CG98, the analogues of eqs. (1) are
\begin{equation} 
\cases{
\ddphi   = -\displaystyle{{\Omegac^2\DItd\over\Iu}\varphi +
            {\Omegac (\Iu -\DItd)\over\Iu}\dtheta}, \cr 
\ddtheta = -\displaystyle{{(\Omegac^2 +\DTut)\DItu\over\Id}\vartheta -
            {\Omegac (\Id -\DItu)\over\Id}\dphi},\cr}
\end{equation}
while the equation for $\psi$ remains unchanged.  The explicit
solution of eqs. (8) is given in Appendix A.  In this case there are
two different equilibrium configurations: the first corresponds to the
galaxy's major axis directed toward the cluster center, and the second
to the galaxy major axis perpendicular to the orbital plane (see
CG98). For simplicity in this paper we restrict to the first case
only.

Due to the system configuration we now need {\it two} coordinate
transformations in order to obtain the equations of motion for a star
in the galaxy. The first transforms the (cluster) inertial system $C$
in the system $\Cc$ centered on the galaxy center of mass (rotating
with the angular velocity $\Omegvc$ and with its $z$ axis parallel to
the $z$ axis of $C$), and the second transforms $\Cc$ in the system
$C'$ in which the galaxy is at rest and its inertia tensor is in
diagonal form. The relation between the position vector $\xv$ in $C$
and the position vector $\csiv$ in $C'$ is given by $\xv=\rv
+\Rotp\csiv$, where $\Rotp=\Rotc\Rot$. $\Rot$ is the same as in
eq. (4), while
\begin{equation}
\Rotc = 
\left(\matrix{
               \cos\Omegac t   &   -\sin\Omegac t   & 0 \cr
               \sin\Omegac t   &    \cos\Omegac t   & 0 \cr
               0               &    0               & 1 \cr
              }\right).
\end{equation}
Straightforward algebra then shows that the equations of motion for a
star as seen from $C'$ are similar to eq. (4), where $\Omegv$ is
substituted by $\Omegvp=\Omegv+\RotT\Omegvc$, $\Omegv$ is given in
eq. (5), and
\begin{eqnarray}
\RotT\Omegvc &=&(\sphi\spsi -\cphi\sthe\cpsi,\nonumber\\ 
             &&\sphi\cpsi +\cphi\sthe\spsi,
               \cphi\cthe)\Omegac. 
\end{eqnarray}
Moreover, $\RotT\ddxv$ is substituted by $\RotpT(\ddxv-\ddrv)\simeq
-\nabla_{\csiv}\phig +\RotT\Tc\Rot\csiv$, where now $\Tc$ is the CTF
tensor as seen in $\Cc$. Note that in tidal approximation, the center
of mass of the galaxy will consistently remains in the circular orbit,
independently of the galaxy oscillations (see CG98). In Sects. 3 and
4, and in Appendices A, B, and C we give the explicit expression for
the various quantities of interest for the adopted galaxy and cluster
models.

\section{Galaxy and cluster models}

We now present the specific galaxy and cluster density distributions
adopted in the numerical simulations, and we derive the required
explicit expressions for the galaxy inertia tensor and the CTF: a
similar approach can be found in Valluri (1993). For simplicity we
assume that the galaxy and cluster densities are stratified on
homeoids. In MC03ab we adopted triaxial $n=0$ Ferrers (1887) and
Hernquist (1990) models, respectively: here we recall the results
obtained when adopting the ellipsoidal generalization of the widely
used $\gamma$-models (Dehnen 1993):
\begin{equation}
\rhog (m) = {\Mg\over \alphau\alphad\alphat}
{3-\gamma\over 4\pi} {1\over m^\gamma (1+m)^{4-\gamma}},\quad
0\leq\gamma <3,
\end{equation} 
while we perform the extended exploration of the parameter space by
using the density distribution
\begin{equation}
\rhog (m) = {\Mg\over \alphau\alphad\alphat}
{15\over 2\pi} {1\over (1+m)^6},
\end{equation} 
where $\Mg$ is the total mass of the galaxy, and
\begin{equation}
m^2 =  \sum_{i=1}^3 {\csici^2\over\alphai^2},
\qquad \alphau\geq\alphad\geq\alphat. 
\end{equation}
Quite obviously, the density profiles in eq. (11) are more realistic
than that in eq. (12): however, for technical reasons described in
Appendix C, this latter density profile allows for much more faster
numerical simulations (i.e., a larger number of
particles). Fortunately, the obtained escape rates in all the explored
models, here and in MC03ab, are remarkably similar, and so can be
considered quite robust (within the considered scenario).

The inertia tensor of generic homeoids are given by
\begin{equation}
\Ii={4\pi\over 3}\alphau\alphad\alphat (\alphaj^2+\alphak^2)\hg,
\end{equation}
where $\hg =\int_0^{\infty}\rhog (m)m^4 dm$, and so the stability
requirement $\Iu\leq\Id\leq\It$ is satisfied.  {\it Note also that,
according to eq. (3), the frequencies for homeoidal stratifications do
not depend on the specific density distribution assumed, but only on
the quantities $(\alphau ,\alphad ,\alphat)$}.  For a better
comparison with observations, in the following we use the two
ellipticities\footnote{In fact, it can be easily proved the for
generic homeoidal distributions, $\epsilon$ and $\eta$ defined
according to eq. (15) are the true ellipticities in the relative
projection planes.}
\begin{equation}
{\alphad\over\alphau}\equiv 1 -\epsilon, \quad\quad
{\alphat\over\alphau}\equiv 1 -\eta,\quad 
\epsilon\leq\eta\leq 1.
\end{equation}

A rough estimate of characteristic stellar orbital times inside $m$
for an homeoidal distribution is given by $\Porb (m)\simeq 4\Pdyn (m)
= \sqrt{3\pi /G\bar{\rhog} (m)}$, where for a $\gamma$ model the
{\it mean} galaxy density inside $m$ is
\begin{equation}
\bar{\rhog} (m) = {3\Mg\over 4\pi\alphau\alphad\alphat}
                  {1\over m^\gamma (1+m)^{3-\gamma}},
\end{equation}
and the galaxy mass inside $m$ is given by:  
\begin{equation}
\Mg (m) = {4\pi\alphau\alphad\alphat}\int_0^m \rhog (t)t^2 \,dt=
\Mg {m^{3-\gamma}\over (1+m)^{3-\gamma}}.
\end{equation}
Thus,
\begin{eqnarray}
\Porb (m)&\simeq& 9.35\times 10^6 
         \sqrt{{\alphaun^3 (1-\epsilon)(1-\eta)\over\Mgn}}\nonumber\\
         &&m^{\gamma/2}(1+m)^{({3-\gamma})/2} \quad {\rm yrs}, 
\end{eqnarray}
where $\Mgn\equiv\Mg/10^{11}\msun$ and $\alphaun\equiv\alphau/{\rm
kpc}$ is the galaxy ``core'' major axis\footnote{We recall here that
for the spherically symmetric Hernquist model, (i.e., $\alphau
=\alphad =\alphat$ in eq. [12]), $\Reff\simeq 1.8 \alphau$, while for
the density distribution in eq. (12) $\Reff\simeq 0.75\alphau$.};
thus, in the outskirts of normal galaxies orbital times well exceeds
$10^8$ or even $10^9$ yrs. A similar conclusion is reached also for
the model in eq. (12).

To describe the cluster density distribution we use the simple mass
profile
\begin{equation}
\rhoc (m) = {\rhocz\over (1+m^2)^2},
\end{equation}    
where $m$ is given by an identity similar to eq. (13), with
$\acu\geq\acd\geq\act$; in the case of the spherical cluster we assume
$\acu =\acd =\act$. In the two following Sections we will compute the
CTF components associated with eq. (19).

\subsection{Galaxy at the cluster center}

In Appendix B it is shown that the tidal field components at the
center of a non-singular homeoidal distribution (such as that in
eq. [19]), are given by
\begin{equation}
\Ti=-2\pi G\rhocz\wci .
\end{equation}
Note that the quantities $\wci$ do not depend on the specific density
profile, and that $\wcu\leq\wcd\leq\wct$ for $\acu\geq\acd\geq\act$,
thus fulfilling the conditions for stable equilibrium in eq. (1). In
analogy with definitions (15), we introduce the two cluster
ellipticities, $\acd/\acu\equiv 1-\mu$ and $\act/\acu\equiv 1-\nu$,
with $\mu\leq\nu\leq 1$, and the expansion of the quantities $\wci$
for small flattenings is given in Appendix C.

In order to determine the galaxy oscillation frequencies we need,
according to eqs. (3) and (20), to have a realistic estimate of
$\rhocz$. Unfortunately this quantity is not well measured in real
clusters, and for its determination we use the virial theorem,
$\Mc\sigvd = -U$ ($U$ is the cluster gravitational potential energy
and $\sigvd$ is the virial velocity dispersion, that we assume to be
estimated by the observed velocity dispersion of galaxies in the
cluster). The explicit calculations of the factor $2\pi G\rhocz$ is
presented in Appendix B.

We now compare the galactic oscillation periods $\Pphi
=2\pi/\omegphi$, $\Pthe =2\pi/\omegthe$, and $\Ppsi =2\pi/\omegpsi$,
with the characteristic stellar orbital times $\Porb$ in galaxies: for
simplicity we give here the expansions for small cluster and galactic
flattenings. From eqs. (3), (14), (20), and using eqs. (C.8)-(C.12) we
finally obtains to the leading order in the flattenings
\begin{equation}
\Pphi \simeq {8.58\times 10^8\over\sqrt{(\nu -\mu)(\eta -\epsilon)}}
             {\acun\over\sigvn}\quad {\rm yrs},
\end{equation}
\begin{equation}
\Pthe \simeq {8.58\times 10^8\over\sqrt{\nu\eta}}
             {\acun\over\sigvn}\quad {\rm yrs},
\end{equation}
\begin{equation}
\Ppsi \simeq {8.58\times 10^8\over\sqrt{\mu\epsilon}}
             {\acun\over\sigvn}\quad {\rm yrs},
\end{equation}
where we normalized $\sigv$ to 1000 km/s and $\acu$ to 250 kpc.  So, a
comparison of quantities above with eq. (18) shows that in the outer
halo of giant Es the stellar orbital times can be of the same order of
magnitude as the oscillatory periods of the galaxies themselves.  For
example, in a relatively small (Hernquist) galaxy of $\Mgn =0.1$ and
$\alphaun =1$, $\Porb\simeq 1$ Gyr at $m\simeq 10$ (i.e., at $\simeq
5\Reff$), while for a galaxy with $\Mgn =1$ and $\alphaun =3$ the same
orbital time is at $m\simeq 7$ (i.e., at $\simeq 3.5\Reff$).

\subsection{Galaxy in circular orbit}

In the case of spherical cluster, the CTF tensor in the reference
system $\Cc$ introduced in Sect. 2.2 is
\begin{equation}
{\bf T}=-\Omegac^2(r)
\left(\matrix{3q-2 & 0 & 0 \cr
                0  & 1 & 0 \cr
                0  & 0 & 1 \cr}
\right),
\end{equation}
where $q(r) \equiv \rhoc (r)/\rhocm (r)$ and $\rhocm\equiv 3\Mc
(r)/4\pi r^3$ (see, e.g., CD94, CG98).  From eq. (19) we obtain
\begin{equation}
\Mc(r)= 2\pi\acu^3\rhocz\left(\arctan\rt-{\rt \over 1+\rt^2}\right), 
\end{equation}
\begin{equation}
\rhocm (r)={3\rhocz\over 2\rt^3}\left(\arctan\rt-{\rt\over 1+\rt^2}\right),
\end{equation}
\begin{equation}
q (r) ={2\rt^3\over 3\left(1+\rt^2\right)^2}
\left(\arctan{\rt}-{\rt\over 1+\rt^2}\right)^{-1},
\end{equation}
where $\rt\equiv r/\acu$.  In analogy with Sect. 3.1, we now compare
the galaxy oscillation periods with the mean stellar orbital times,
and for simplicity we restrict to the special case of the galaxy
oscillating around the $\xi_3$-axis only, i.e., we fix $\varphi =0$
and $\vartheta =0$ in eq. (10). Thus, from the results of Sect. 2.2,
and from eq. (24)
\begin{equation}
\Ppsi\equiv {2\pi\over\omegpsi}\simeq
            {2\pi\over\Omegac(r)}
            {1\over\sqrt{3\epsilon [1-q(r)]}},
\end{equation}
where we used the last of eqs. (3) and the expansion for small
flattenings of the coefficient involving the galaxy moments of
inertia. In Fig. 1 we plot $\Ppsi$ as a function of $r$ for the two
spherical cluster models used in the simulations described in Sect.
5.2, in the cases of an E2 and and E4 galaxy; note that $\Ppsi$
decreases at increasing galaxy flattening, while its radial trend is
non monotonic, with a minimum at $r\simeq a_1$, the cluster core
radius. The peculiarity of this location has been already noted (e.g.,
see CD94), and its impact on orbital evolution will be evident in
Sect. 5.2.
\begin{figure}[htbp]
\psfig{file=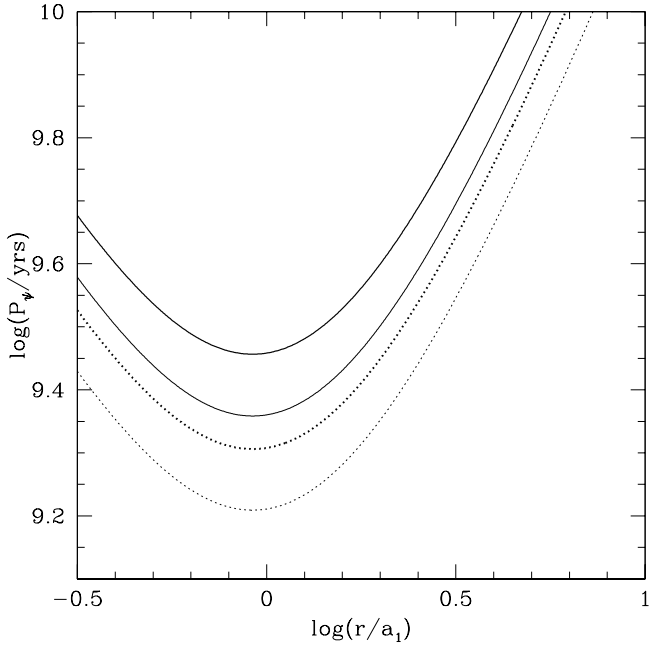,width=9cm,height=9cm,angle=0}
\caption[]{The oscillation period around the $\xi_3\parallel\Omegvc$ axis for 
a galaxy in circular orbit of radius $r$ in a cluster of core radius
$a_1$ and velocity dispersion $\sigv$. Solid lines refer to an E2
galaxy, while dotted lines to an E4 galaxy. Cluster model Ca (heavy
lines) is characterized by $a_1=250$ kpc and $\sigv=1000$ km/s, while
model Cb by $a_1=100$ kpc and $\sigv=500$ km/s.}
\label{fig1}
\end{figure}
\section{The numerical integration scheme}

\subsection{Dimensionless equations}

In the numerical simulations we use as time normalization the Hubble
time $\tH =15$ Gyrs, as mass normalization the galaxy mass $\Mg$, and
as length normalization $\alphau$, the homeoidal scale-length
corresponding to the galaxy major axis. Accordingly, eq. (4) in
dimensionless form reads
\begin{eqnarray}
{d^2\tcsiv\over d\tau^2}&=&
-{G\Mg\tH^2\over\alphau^3}\nabla _{\tcsiv}\tphig\nonumber\\
                        &&-2\omegac\tH\tOmegv\wedge\tnuv
                          -\omegac^2\tH^2 [
                                    \dtOmegv\wedge\tcsiv +
                                    \tOmegv\wedge (\tOmegv\wedge\tcsiv)],
\end{eqnarray}
where $\tau\equiv t/\tH$ and $\phig\equiv (G\Mg/\alphau)\tphig$; note
that from eq. (3) $\omegi =\omegac\tomegi$ where $\omegac\equiv \sqrt
{2\pi G\rhocz}$, and so from eqs. (2) and (5)
$\Omegv=\omegac\tOmegv$. In addition, consistently with the adopted
``small oscillations approximation'', the angular velocity is used in
linearized form
\begin{equation}
\Omegv\simeq (\dphi, \dtheta, \dpsi),
\end{equation}
i.e., in eq. (5) we retained only linear terms in $\phiM$, $\theM$,
$\psiM$; note that we neglect the direct effect of the cluster
gravitational field on stellar orbits (cf. with eq. [6]).


According to the discussion in Sect. 2.2, the dimensionless
equations of the motion for a star in the galaxy on a circular orbit
are similar to eq. (29), where now $\omegac\equiv \sqrt {2\pi
G\rhocm}$, we again disregard the direct effect of the CTF, $\tOmegv$
is substituted by $\tOmegvp\equiv \Omegvp/\omegac$, and to the linear
order
\begin{equation}
\Omegvp\simeq
(\dphi -\vartheta\Omegac,\dtheta +\varphi\Omegac,\dpsi+\Omegac).
\end{equation}
Note that $\Omegac/\omegac=\sqrt{2/3}$.  

\subsection{Initial conditions}

To integrate the second order differential equations (29) describing
the motion of a test star in the non-inertial reference frame fixed in
the galaxy, we use a code based on an adapted f77, double-precision
4th order Runge-Kutta routine (Press et al. 1986), with adaptive time
step.  In practice, for a given galaxy and cluster model, we analyzed
the behaviour of a large number of orbits ($\Ntot\sim 10^4\div 10^6$)
over an equivalent Hubble time: from this point of view our
simulations are nothing else that a large collection of one-body
problems in time dependent potentials. Of course, the initial
conditions for each orbit must be realized with some care.  First of
all, in order to properly sample the galaxy density profile the
initial positions are obtained from the Von Neumman {\it rejection
method} (Press et al. 1986) which uses the (elliptical) cumulative
mass distribution $\Mg (m)/\Mg$ as probability function for the
homeoidal radius; the two position angles are then extracted from
uniformly populated standard angular intervals (i.e., $0\leq\varphi <
2\pi$ and $-1\leq\cos\vartheta\leq 1$).

A more delicate issue is represented by the assignment of the three
components of the initial velocity. In principle, one should perform a
von Neumann rejection method in phase-space, provided the model
phase-space distribution function is known at each point.
Unfortunately, for triaxial galaxies the only sufficiently general
method is the Schwarzschild (1979) orbit linear superposition method,
which however for our purposes would be very demanding from the
numerical point of view. Here we adopt a more empirical approach,
whose consistency is however readily verified. In some simulations,
for each initial position in configuration space, we calculate the
1-dimensional velocity dispersion $\sigma^2$ from the Jeans equations
of the {\it spherical} analogue of the density distribution in
eqs. (11) and (12), we choose a dimensionless parameter $\chi\leq 1$,
and we randomly distribute over the three directions the initial
particle velocity so that its square modulus equals
$3\chi\sigma^2$. In other simulations, we same procedure was repeated,
but assuming a fixed (within a given simulation) fraction $\chi$ of
the local escape velocity $v_{\rm esc}$.

For each simulation (i.e., for assigned galaxy and cluster models,
maximum oscillation angles, and $\chi$ value) we actually run the code
{\it twice}: in the first simulation the galaxy maximum oscillation
angles are set to zero (in other words the galaxy is at rest at the
equilibrium position in the CTF), and all the initial conditions
leading to unacceptable orbits for the unperturbed case are discarded.
In this way we can also test the code for energy conservation.  In all
the computed models the ``unperturbed'' case pass very well the
acceptability test, with relative errors on energy conservation for
each orbit never exceeding $\sim 10^{-5}$ over an Hubble time, and
without significant spurious evaporation. Thus, we can safely assume
that the evaporation rates found in the explored models are a genuine
product of galaxy oscillations, and no a product of and inconsistent
assignment of the initial conditions.

All the simulations have been performed on GRAVITOR, the Geneva
Observatory 132 processors Beowulf cluster\footnote{For technical
detail on GRAVITOR see
http://obswww.unige.ch$/~$pfennige/gravitor/gravitor$\_$e.html}: for
reference, the computation of $10^4$ orbits requires $\sim 2$ hours
when using 10 nodes.

\section{Results}

\subsection{Galaxy at the center of a triaxial cluster}

Here we describe the results of the first of the two cases discussed
in Sect. 2, namely the case of a triaxial galaxy at the center of a
triaxial cluster. In this case we already have preliminary
informations for galaxy models described by Ferrers (1887, see MC03a)
and by the more realistic Hernquist (see MC03b) density
distribution. In this Sect. we complement these informations with the
results obtained when using the density profile in eq. (12). However
in the present case, as for Hernquist models and at variance with
Ferrers distribution, the concept of ``escape'' is quite subtle. In
fact, while in truncated galaxy models the escape criterion is
obvious, in untruncated models the ``escaping'' particles must be
identified in a different way. In particular we choose to plot, for
each test particle in a given model, the ratio $\mmax/\mi$ vs. $\mi$,
where $\mi$ is the ellipsoidal radius of a given initial condition,
while $\mmax$ is the maximum ellipsoidal distance reached by that
particle over an equivalent Hubble time. We present here only a
representative simulation among the several performed, because they
remarkably confirm the previous results for Hernquist models described
in MC03b. In particular, the shown simulation refers to a galaxy model
characterized by the two flattenings $\epsilon = 2/5$ and $\eta =
16/25$; the same ellipticities have been also adopted for the cluster,
i.e. $\mu = 2/5$ and $\nu = 16/25$. Thus, the adopted galaxy model
would be classified, as a function of the observation angle, from an
$E4$ up to an $E6$ galaxy. Its mass is fixed to $\Mg = 10^{11}\msun$,
while its core semimajor axis is fixed to $\alphau=6.58$ kpc
(corresponding to an effective radius in the spherical model of
$\simeq 5$ kpc).  The cluster is characterized by $\sigv = 1000$ km/s
and $\acu =250$ kpc. Initial conditions were generated with $\mi\leq
8$ and with a square modulus of the initial velocity of each particle
equals to the local $0.3v_{\rm esc}^2$.  The galaxy maximum
oscillation angles along the three directions are
$\phiM=\theM=\psiM=0.25$.

\begin{figure}[htbp]
\psfig{file=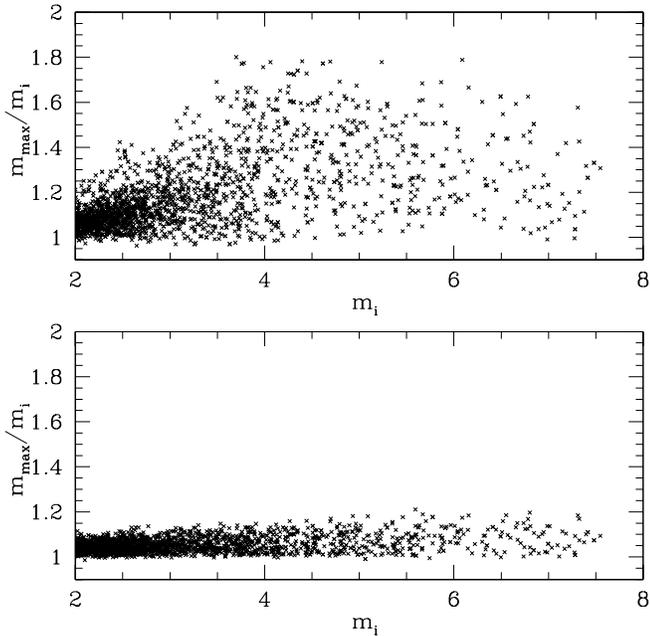,width=9cm,height=9cm,angle=0}
\caption[]{The ratio $\mmax/\mi$ vs. $\mi$ for the subset of particles with 
$\mi\geq 2$, evaluated over $\tH$. Upper panel: unperturbed galaxy with 
$\phiM=\theM=\psiM=0$. Lower panel: the same model with 
$\phiM=\theM=\psiM=0.25$ rad.}
\label{fig2}
\end{figure}

The results are shown in Fig.~\ref{fig2}. Note how in the unperturbed
case, with the adopted $\chi$ value, $\mmax/\mi$ increases only
slightly, while in the oscillating case this number reaches higher
values, of the order of 1.5 or more. These results are very similar to
those described in MC03b (see Figs. 1 and 2 there). Note also how the
expansion effect is more important in the external galaxy regions
(affecting particles with $\mi\gsim 3$), even though we recall that in
the simulations the direct effect of the CTF is {\it not} considered:
as a consequence, the obtained galaxy expansion is a genuine effect of
galactic oscillations. By the way, this is a quite interesting result,
that one could relate to the process of cD formation. In fact, it is
well known that it is not easy to grow with galactic cannibalism, at
least in numerical simulations, the extended cD halos (e.g., see
Nipoti et al. 2003, and references therein). However, one should
recall that the CTF in shallow cores is compressive (e.g., see Valluri
1993, CD94): as a consequence, the final galaxy density profile will
be determined by the competing effect of galaxy oscillations and the
cluster gravitational field. Thus, in order to quantitatively discuss
the possibility that the galaxy sloshing at the cluster center is at
the origin of the cD halos, high-resolution, self-consistent N-body
simulations are required.

\subsection{Galaxy on circular orbit in a spherical cluster}

We now move to describe the results of the simulations of galaxies in
circular orbit around the cluster center: altough still highly
idealized, this cases apply to a larger number of galaxies than the
previous one. At variance with the case of a galaxy at the cluster
center, galaxies in circular orbit were not studied before, so here we
present a larger set of simulations, organized in two groups. In the
first set of simulations the spherical cluster is characterized by
$\sigv =1000$ km s$^{-1}$ and $a_1 = 250$ kpc (model Ca), while in the
second set $\sigv = 500$ km s$^{-1}$ and $a_1 = 100$ kpc (model
Cb). For each model cluster we study {\it three} different positions
of the galaxy center of mass, namely circular orbits with $r/a_1 =0.5$
(models Cai and Cbi), with $r/a_1 =1$ (models Cac and Cbc), and
finally with $r/a_1 =2$ (models Cao and Cbo). In order to reduce the
dimensionality of the parameter space, in all the simulations we fix a
galaxy model with $\Mg =10^{11}\msun$, $\alpha_1=7$ kpc, $\epsilon
=0.2$ and $\eta =0.4$ (thus corresponding to a less flattened galaxy
than that used in Sect. 5.1). Finally, we explore different cases of
initial velocities of the test particles, with $\chi$ in the range 0.1
- 0.5. In all cases, we restrict our analysis to the case of a galaxy
oscillating only around the $\xi_3$ axis (the axis perpendicular to
the orbital plane of the galaxy center of mass), and we adopt $\psiM
=0.2$ rad.

\begin{figure*}[htbp]
\centering
\psfig{file=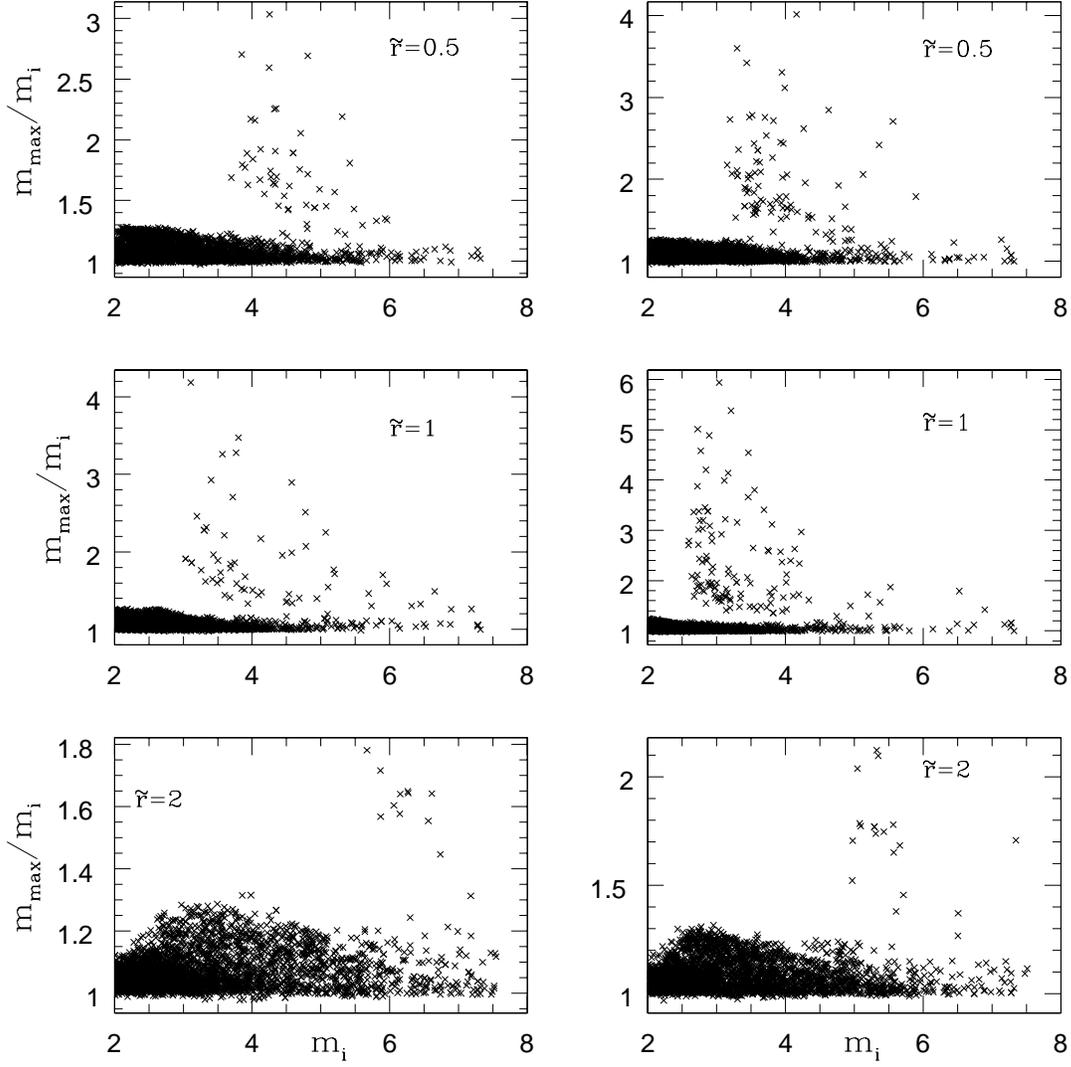,width=15cm,height=15cm,angle=0}
\caption[]{The ratio $\mmax/\mi$ vs. $\mi$ for the subset of particles with 
$\mi\geq 2$, evaluated over $\tH$, in the case of a galaxies in 
circular orbit around the cluster center. {\it Left panels}: from the 
top, models Cai, Cac, and Cao, respectively. {\it Right panels}: from the 
top, models Cbi, Cbc, and Cbo, respectively. In all cases the 
maximum oscillation angle is set to $\psiM=0.2$ rad.}
\label{fig3}
\end{figure*}
We note here that the set-up of the initial conditions is considerably
more time-expensive that in the case of the galaxy at the cluster
center. In fact, in the present case, ``galaxy at rest'' means that
the galaxy center of mass is actually (uniformly) {\it rotating}
around the cluster center, and this induces a centrifugal acceleration
on the galaxy stars, as apparent by setting
$\varphi=\vartheta=\dot\psi=0$ in eq. (31). Thus, if one erroneously
adopt as stellar orbit initial conditions the same conditions adopted
in the previous Section, the result is a vigorous stellar escape, just
due to the centrifugal acceleration. Accordingly, test numerical
simulations (not shown here) were characterized by a significant
number of stars escaped up to $\mmax\simeq 100$. Of course this is not
a physical process, but only the result of the non-equilibrium initial
conditions: in order to avoid this problem we proceeded as
follows. For each galaxy model, we first performed the orbital
calculation for a large number (say $10^5$) initial conditions
arranged as in Sect.  5.1, maintaining the galaxy in uniform rotation
around the cluster center and without oscillations. We then discarded
all the initial conditions corresponding to orbits for which, over an
Hubble time, $\mmax/\mi\gsim 1.2$, in order to mimic the result shown
in the bottom panel of Fig. 2. We finally used the remaining initial
conditions to study the orbital evolution in the corresponding {\it
oscillating} galaxy. The results are summarized in Fig. 3.

Several interesting comments can be made from inspection of
Fig. 3. The first, and more obvious, is that also in the rotating
galaxies, oscillations are expected to produce stellar
evaporation. However, in the rotating cases we invariably found that
the largest $\mmax/\mi$ are larger that the corresponding quantity
relative to the galaxies at the cluster center (cfr. Fig. 2), and this
is due to the additional effect of the centrifugal term on escaping
stars. Strictly related to this point is also the systematic behavior
of $\mmax/\mi$ shown in Fig. 3, where it is apparent how the largest
ratios {\it systematically decrease} for galaxy models placed at
$r=\acu$, $\acu/2$ and $2\acu$, respectively. This non-monotonic trend
si just a reflection of the same behavior of $\Ppsi$, shown in Fig. 1:
shorter the galaxy oscillation times, larger are the $\mmax/\mi$ that
can be reached by escaping stars. In this respect, the peculiar
dynamical situation of galaxies orbiting near the cluster core radius
it is well known: not only there is a change in the topological nature
of the CTF there (see, e.g., CD94), but also oscillation periods are
the shortest (for a given galaxy model). The third comment about
Fig. 3 is the common pattern of the radial behavior of the upper
envelope of $\mmax/\mi$, i.e., a quite well defined, sharp increase
followed by a somewhat smoother decline at increasing $\mi$. This
behavior basically corresponds to the place in the galaxy where
orbital times (see eq. [18]) become comparable with the oscillation
times. Inside this radius stellar orbits react adiabatically to the
forcing due to galaxy oscillations (see CD94), while outside galaxy
oscillations become effective in changing significantly the stellar
orbits. As expected, the effect becomes however smaller and smaller
for stellar orbits with increasing orbital times. Finally, note how
there are not significant differences between the results in the large
(Ca) and small (Cb) clusters: we interpret this as another
manifestation of the comparable oscillatory periods in the two cases
(see Fig. 1).


\section{Discussion and conclusions}

In this paper we presented a representative selection of numerical
models aimed at the study of collisionless evaporation of stars from
cluster elliptical galaxies. From a theoretical point of view this
possibility was pointed out in CG98, who showed that characteristic
oscillation times of elliptical galaxies in near equilibrium
configurations in the tidal field of the parent cluster are curiously
comparable to the stellar orbital times in the outskirts of the
galaxies themselves. From a numerical point of view the present work
represents the natural follow-up of two recent preparatory works on
the same subject (MC03ab), where we explored the simplest equilibrium
configuration (i.e., a galaxy with the center of mass at rest at the
center of a triaxial cluster), but we left unaddressed the more
complicate (although more astrophysically representative) case of
galaxies with their center of mass in rotation around the (spherical)
cluster center. We recall that in all the presented simulations the
{\it direct} effect of the cluster gravitational field on stellar
orbits was {\it not} included, in order to study the effects of galaxy
oscillations only; we also recall that its effect is {\it compressive}
inside the cluster core radius, while {\it expanding} outside (see,
e.g., Valluri 1993, CD94).

The main results can be summarized as follows:

\begin{itemize}

\item A general analytical framework suitable for the study of orbital
evolution of stars in galaxies oscillating around their stable
equilibrium positions in the tidal field of the parent cluster is
presented. In particular, we analyzed the cases of triaxial galaxies
at the center of a triaxial cluster and in circular orbit around the
center of a spherical cluster.

\item It is shown that, for realistic galaxy and cluster models,
galaxy oscillation times can be comparable to characteristic stellar
orbital times in the external galactic regions, and thus important
effects on stellar orbits are expected there.

\item Numerical simulations of galaxiesat the cluster center
remarkably confirm the preliminary results presented in MC03ab (where
different galaxy models were adopted). In particular, we found that
orbits in the external parts of the parent galaxy are affected by
galaxy oscillations, with stars outside $\mi\simeq 3\simeq 2.25\Reff$
nearly doubling the apocenters. {\it Globally, a galaxy mass fraction
of the order of 10\% is affected by this process}.

\item We speculate that the findings above could be related to the
formation of cD galaxies. In fact, it is well known that numerical
simulations of galactic cannibalism, although successful at
reproducing BCGs, are in general unable to obtain the extended halos
of cDs. In this respect, the fact that the external parts of the
galaxies are the only ones affected by collisionless evaporation, is a
nice property of the explored scenario. Unfortunately, our simulations
cannot be used to study the resulting galactic profiles, and
high-resolution N-body simulations should be performed to study
further the matter.

\item For galaxies in circular orbits, the effect of collisionless
evaporation is enhanced with respect to the previous case, and stars
can reach significantly larger distances. This is not due to the
direct effect of the CTF, but to galaxy oscillations only. In
accordance with our preparatory analysis, we found that the radial
trend of galactic oscillation periods at increasing distance from the
cluster center is reflected by the location in the galaxy of the
expanded region, being the most affected galaxies those orbiting near
the cluster core radius. Even in this cases, {\it the affected galaxy
mass fraction ranges is of the order of 5\% - 10\%, in interesting
agreement with observational estimates}.

\item The natural consequence of the results listed above is that, if
collisionless evaporation is of importance in the production of the
ISP (as direct contributor and/or help to the stripping phenomena due
the direct CTF effect and to galaxy-galaxy encounters, as investigated
in detail by Merritt [1983, 1984, 1985]), the density of ISP (when
normalized to the local cluster luminosity in galaxies) should be
enhanced around the cluster core radius, and rapidly declining at the
cluster outskirts, due to the increase of the galaxy oscillation
periods with the distance from the cluster center.

\end{itemize}

We finally conclude by pointing out that both orbital cases explored
in this paper are rather exceptional.  First of all, we only
considered galaxy models that were initially not tumbling nor
rotating. In addition, most cluster galaxies neither rest in the
cluster center nor move on circular orbits, but they move on elongated
orbits with very different pericentric and apocentric distances from
the cluster's center; in a triaxial cluster many orbits are boxes and
some orbits can be chaotic. These latter cases can be properly
investigated only by direct numerical simulation of the stellar
motions inside the galaxies, coupled with the numerical integration of
the equations of the motion of the galaxies themselves. In addition, a
statistically significant galaxy population made by the sum of
galaxies with different masses, scale-lengths, and flattenings, should
be considered.

\section{Acknowledgements}

We would like to thank Magda Arnaboldi, Giuseppe Bertin, Daniel
Pfenniger for useful discussions, and the anonymous Referee for
helpful comments.  V.M. is grateful to Geneva Observatory for allowing
the use of their GRAVITOR beowulf cluster.

\appendix

\section{Oscillation angles for a triaxial galaxy in circular orbit}

Here we give the explicit solution of eqs. (8) for assigned initial
conditions $(\varphi_0,\dphi_0)$ and $(\vartheta_0,\dtheta_0)$. As
shown in CG98, the two mixed equations of the second order 
\begin{equation}
\cases{\ddphi   =-A_1\varphi + B_1\dtheta,\cr
       \ddtheta =-A_2\vartheta -B_2\dphi,
}
\end{equation}
(where the 4 coefficients $A_1,A_2,B_1,B_2$ are given in eqs. [8]),
can be separated in two identical biquadratic equations, each of them
involving only one of the two variables $\varphi$ and $\vartheta$.
For example, the equation for $\varphi$ can be written as
\begin{equation}
{d^4\varphi\over dt^4}+A{d^2\varphi\over dt^2}+B=0,
\end{equation}
where
\begin{eqnarray}
A&\equiv& A_1+B_1B_2+A_2=\Omegac^2\times\nonumber\\
 &&\left[{(u+v-1)^2\over uv}+{1-v\over u}+{(4-3q)(1-u)\over v}\right],
\end{eqnarray}
\begin{equation}
B\equiv A_1A_2=\Omegac^4\times 
        {(4-3q)(1-u)(1-v)\over uv},
\end{equation}
$u\equiv\Iu/\It$, $v\equiv\Id/\It$, and the functions $\Omegac(r)$ and
$q(r)$ are given in Sect. 3.2. The standard substitution $\varphi =
X \exp(i\omega t)$ in eq. (A.2) leads to the characteristic equation
$\omega^4 -A\omega^2 + B=0$, whose solutions can be written as
\begin{equation}
\omegapm\equiv\sqrt{\lambdapm};\quad\lambdapm = {A\pm\sqrt{A^2 -4B}\over 2};
\end{equation}
note that in case of a stable equilibrium for the galaxy configuration
(as assumed in this paper), the roots $\lambdapm$ are positive, and
that from eqs. (A.3)-(A.4) $\omega\propto\Omegac$. Thus, the general
solution of eqs. (8) can be written as
\begin{equation}
\cases{\varphi   =a\cos(\omegam t + a_0)+b\cos(\omegap t +b_0),\cr
       \vartheta =c\cos(\omegam t + c_0)+d\cos(\omegap t +d_0),
}
\end{equation}
where the four amplitudes $a,b,c,d$ and phases $a_0,b_0,c_0,d_0$ must
be determined by imposing the 4 initial conditions at $t=0$. The
remaining four constraints are provided by the request that eqs. (A.6)
satisfy eqs. (A.1) at any time:
\begin{equation}
\cases{
c_0=a_0+\pi/2,\cr 
d_0=b_0+\pi/2;\cr
a(\omegam^2 - A)-c B_1\omegam=0,\cr 
b(\omegap^2 - A)-d B_1\omegap=0:
}
\end{equation}
the identities involving amplitudes can be also written in terms of
$C$ and $B_2$ instead of $A$ and $B_1$. We can now proceed to impose
the initial conditions on eqs. (A.6). The easiest way is to define the
6 new quantities
\begin{equation}
\cases{
\rom\equiv a/c = B_1\omegam/(\omegam^2 -A),\cr
\rop\equiv b/d = B_1\omegap/(\omegap^2 -A),\cr
X=\cos a_0,\quad Y=\cos b_0,\cr       
U=\sin a_0,\quad V=\sin b_0:
}
\end{equation}
note that $\rop$ and $\rom$ are known. Evaluating eqs. (A.6) at $t=0$,
and using eqs. (A.7)-(A.8) one obtains
\begin{equation}
\cases{\rom cX + \rop dY = \varphi_0,\quad
       \omegam cX + \omegap dY = -\dtheta_0,\cr
       c U + d V = -\vartheta_0,\quad
       \omegam\rom cU + \omegap\rop dV = -\dphi_0,\cr
       X^2+U^2=1,\quad Y^2+V^2=1.
}
\end{equation}
The first 4 equations above can be solved for the unknowns $cX$, $dY$,
$cU$, $dV$ in terms of $r_{\pm}$, $(\vartheta_0,\dtheta_0)$,
$(\varphi,\dphi_0)$: then the two independent phases and amplitudes
are obtained as
\begin{equation}
\cases{c^2=(cX)^2 + (cU)^2,\quad d^2=(dY)^2+(dV)^2,\cr
       \tan a_0 =cU/cX,\quad\tan b_0=dV/dY.
}
\end{equation}
The remaining quantities are finally determined from eqs. (A.7)-(A.9).

\section{Potential, gravitational energy an TF components in homeoidal 
systems} 
As in eqs. (11), (12) and (19) let use assume that
\begin{equation}
\rho =\rhon\rhot (m),
\end{equation}
where $m$ is defined as in eq. (13), and $\rhon$ is a normalization
constant for density. The total mass of the model (when finite) is
given by
\begin{equation}
M = 4\pi\acu\acd\act\rhon\M,
\end{equation}
where 
\begin{equation}
\M\equiv \int_0^{\infty}\rhot (m) m^2dm.
\end{equation} 
For example, in eqs. (11), (12), and (19) $\M = 1/(3-\gamma)$, $\M
=1/30$, and $\M =\pi /4$, respectively. From potential theory, the
general expression for $\phig (\csiv)$ is found using the well-known
formulas (Kellog 1953, Chendrasekhar 1969, BT87),
\begin{equation}
\phi(\xv) = -G\pi\rhon\acu\acd\act \int_0^{\infty}{\DePsit [m(\tau,\xv)] \over
             \Dtau}\, d\tau
\end{equation}
where 
\begin{equation}
\Dtau = \sqrt{(\acu^2+\tau)(\acd^2+\tau)(\act^2+\tau)},
\end{equation}
\begin{equation}
\DePsit (m) = 2\int_{m(\tau,\xv)}^{\infty} \rhot (t) tdt ,
\end{equation}
and
\begin{equation}
m^2(\tau,\xv) = \sum_{i=1}^3 {\xci^2 \over \aci^2+\tau}.
\end{equation}

For homeoidal density distributions the potential energy tensor
components are given by (Roberts, 1962)
\begin{equation}
U_{ii} = -G\pi^2\rhon^2 \acu\acd\act\aci^2\wci
          \int_0^{\infty}\DePsit^2 (m)dm,
\end{equation}  
where 
\begin{equation}  
\wci =\acu\acd\act\int_0^{\infty}{d\tau \over (\aci^2+\tau)\Dtau},
\end{equation} 
and $U=U_{11}+U_{22}+U_{33}$.  The three integrals $\wci$ are
calculated by using identities 238.03, 238.04 and 238.05 from Byrd \&
Friedman (1971):
\begin{equation}
\wcu= 2\tacd\tact {F(\varphi ,k)-E(\varphi ,k)\over k^2\sin^3\varphi},
\end{equation}
\begin{equation}
\wcd = 2\tacd\tact {E(\varphi,k)-k'^2 F(\varphi ,k)-k^2(\act/\acd)
       \sin\varphi\over k^2 k'^2\sin^3\varphi},
\end{equation}
\begin{equation}
\wct = 2\tacd\tact{(\acd/\act)\sin\varphi-E(\varphi ,k)\over 
       k'^2\sin^3\varphi}.
\end{equation}
where $\tacd =\acd/\acu$, $\tact =\act/\acu$, $\varphi
\equiv\arcsin\sqrt{1-\tact^2}$, $k^2 \equiv (1-\tacd^2)/(1-\tact^2)$,
and finally
\begin{equation}
\cases{
F(\varphi,k)=\displaystyle{
             \int_0^{\varphi} {dt\over\sqrt{1-k^2\sin^2t}}},\cr 
E(\varphi,k)=\displaystyle{\int_0^{\varphi} \sqrt {1-k^2\sin^2t}\,dt,}
}
\end{equation}
are the Elliptic Integrals of the first and second kind, respectively
and $k'^2=1-k^2$. Also, we recall that the TF tensor is given by
$\Tij=-\partial^2\phi/\partial\xci \partial\xcj$: the explicit
evaluation of this quantity by using eq. (B.4) proves eq. (20). We
also evaluate the quantity $2\pi G\rhocz$ needed there by using the
virial theorem, $\Mc\sigvd = -U$. From eqs. (B.2) and (B.8) we obtain
the exact expression
\begin{equation}
2\pi G\rhon = {8\M \sigvd\over \sum_{i=1}^3\aci^2\wci
          \int_0^{\infty}\DePsit^2 (m)dm}.
\end{equation}
For the density distribution in eq. (19) it can be proved that the
value of the integral at denominator equals $\M =\pi/4$.

\section{Expansions for small flattenings}

The explicit form of the potential $\phi$ generated by a generic
homeoidal density distributions usually cannot be expressed in
explicit, closed form.  On the contrary, its expression in the limit
of small flattenings is trivial. As described in Sect. 3, we
introduce the two ellipticities with $\acd/\acu = 1-\epsilon$ and
$\act/\acu = 1-\eta$; obviously, in spherical models $\epsilon =\eta
=0$.  We now expand eq. (B.4) for small flattenings, retaining only
linear terms.  We perform this expansion at {\it fixed} mass, i.e, we
eliminate the coefficient $\pi\acu\acd\act\rhon$ between eqs. (B.3)
and (B.4).  With the normalization of lengths to $\acu$, a tedious but
straightforward calculation shows that $\phi = -(GM/\acu)
\tilde{\phi}$, where
\begin{eqnarray}
\M\times\tilde\phi &=&{I_2(\rt)\over\rt} + 
                {(3+\epsilon+\eta) [I_1(\infty)-I_1(\rt)]\over 3} +\nonumber\\ 
             &&{(\epsilon+\eta)I_4(\rt)\over 3\rt^3} -
               {(\yt^2\epsilon+\zt^2\eta)I_4(\rt)\over \rt^5},
\end{eqnarray}
and $I_n (x)\equiv \int_0^x\rhot (t)t^ndt$. Many dynamical properties
of general density-potential pairs obtained with this truncation
procedure are presented elsewhere (for a full discussion of this
technique see Ciotti \& Bertin 2004).

For example, for ellipsoidal Hernquist models,
\begin{equation}
I_1(\rt) = {\rt (2+\rt)\over 2 (1+\rt)^2},
\end{equation}
\begin{equation}
I_2(\rt) =  {\rt^2\over 2 (1+\rt)^2},
\end{equation}
\begin{equation}
I_4(\rt) = {\rt (6+9\rt + 2\rt^2)\over 2 (1+\rt )^2}-3\ln (1+\rt).
\end{equation}
Note that, at the center of the Hernquist model, the force remains
finite but is not continuous, i.e., it depends on the approaching
direction, and this fact produces undesidered orbital scattering due
to amplification of numerical errors. This problem is avoided when using a density such that in eq. (12), from which
\begin{equation}
I_1(\rt) = {\rt^2 (10+10\rt +5\rt^2+\rt^3)\over 20(1+\rt)^5},
\end{equation}
\begin{equation}
I_2(\rt) = {\rt^3 (10+5\rt +\rt^2)\over 30(1+\rt)^5},
\end{equation}
\begin{equation}
I_4(\rt) = {\rt^5 \over 5(1+\rt)^5},
\end{equation}
Note that the force now vanishes at the model center.

We also expand up to linear terms the three functions in
eqs. (B.10)-(B.12) for $\mu\to 0$ and $\nu\to 0$:
\begin{equation}
\wcu\simeq 2(1-\mu)(1-\nu)\left ({1\over 3} + {\mu+\nu\over 5}\right ),
\end{equation}
\begin{equation}
\wcd\simeq 2(1-\mu)(1-\nu)\left ({1\over 3} + {3\mu+\nu\over 5}\right ),
\end{equation}
\begin{equation}
\wct\simeq 2(1-\mu)(1-\nu)\left ({1\over 3} + {\mu+3\nu\over 5}\right ).
\end{equation}
With the aid of these expansions we can finally obtain also the
linearization of the gravitational potential energy given in eq. (B.8)
\begin{equation}
U\simeq -{G M^2\over \acu}{\int_0^{\infty}\DePsit^2 (m)dm\over 8\M^2}
\left(1 +{\mu+\nu\over 3}\right).
\end{equation}
and eq. (B.14):
\begin{equation}
2\pi G\rhocz\simeq {4\M \sigvd\over\acu^2 \int_0^{\infty}\DePsit^2
(m)dm} \left [ 1 +{2(\mu +\nu)\over 3}\right ].
\end{equation}

\end{document}